\documentclass[aps,prl,preprint,superscriptaddress]{revtex4}

\usepackage{graphicx}
\usepackage{dcolumn}
\usepackage{bm}
\usepackage[usenames]{color}
\usepackage{colortbl}

\begin{document}

\def\btt#1{{\tt$\backslash$#1}}
\def\BibTeX{\rm B{\sc ib}\TeX}
\draft

\title{Investigation of heavy ions diffusion under the influence 
of current-driven mechanism and compositional waves in plasma}
\author{Vadim Urpin}
\affiliation{A.F.Ioffe Institute of Physics and Technology, 
           194021 St. Petersburg, Russia}

\date{\today}
\begin{abstract}
We consider diffusion caused by a combined influence of the Hall
effect and electric currents, and argue that such diffusion forms 
chemical inhomogeneities in plasma. The considered mechanism can 
be responsible for the formation of element spots in laboratory and 
astrophysical plasmas. Such current-driven diffusion can be 
accompanied by the propagation of a paticular type of waves 
which have not been considered earlier. In these
waves, the impurity number density oscillates alone and their
frequency is determined by the electric currents and sort of 
impurity ions. These compositional waves exist if the magnetic 
pressure in plasma is much greater than the gas pressure. Such 
waves lead to local variations of chemical composition in plasma 
and, hence, can manifest themself by variations of the emission 
in spectral lines.    
\end{abstract}


\maketitle

\section{Introduction}

Often laboratory and astrophysical plasmas are multicomponent, and
diffusion of elements plays an important role in many phenomena. For 
instance, diffusion can be responsible for the formation of chemical 
inhomogeneities which influence emission, heat transport, conductivity, 
etc (\cite{vek87,ren08,kag12}). In thermonuclear fusion 
experiments, the source of impurities is usually the chamber walls, 
and diffusion determines the penetration depth of these impurities and 
their distribution in plasma (\cite{los07,ott11,mol14}). 
Even a small admixture of heavy ions increases drastically radiative 
losses of plasma and changes its thermal properties. In 
astrophysical conditions, chemical inhomogeneities have been 
detected in many stars which have relatively quiescent surface 
layers. It is widely accepted opinion that these inhomogeneities
are determined by diffusion processes \cite{kh85} but,
however, the mechanisms resulting in formation of chemical spots is 
rather uncertain. Usually, diffusion in astrophysical bodies is 
influenced by a number of factors (gravity, radiative force, 
magnetic field, temperature gradient, etc.\cite{spit78}) and, 
therefore, chemically peculiar stars are excellent laboratories to 
study diffusion processes in plasma.

Diffusion in plasma can differ qualitatively from that in neutral 
gases because of the presence of electrons and electric currents. 
This particularly concerns hydrogen plasma where the rate of 
momentum exchange between electrons and protons is comparable to 
the rate of the momentum redistribution among protons \cite{brag65}. 
In such plasma, the influence of electrons on diffusion of heavy 
ions is especially pronounced. Chemical inhomogeneities can appear 
in plasma because of a number
of reasons, for instance, because of a non-uniform temperature.
Also, it is often thought that chemical spots occur due to the 
presence of the magnetic field. The magnetic field can magnetize 
electrons in plasma that, generally, leads to anisotropic 
transport and can produce an inhomogeneous distribution of heavy
ions. Anisotropy of diffusion is characterized by the Hall 
parameter, $x_e = \omega_{Be} \tau_e$, where $\omega_{Be} = e B/m_e c$ 
is the gyrofrequency of electrons and $\tau_e $ is their relaxation 
time; $B$ is the magnetic field. In hydrogen plasma, $\tau_e = 3 
\sqrt{m_e} (k_b T)^{3/2}/4 \sqrt{2 \pi} e^4 n \Lambda $  
\cite{spit78} where $n$ and $T$ are the number density of 
electrons and their temperature, respectively, $\Lambda$ is the 
Coulomb logarithm. At $x_e \geq 1$, the rates of diffusion along 
and across the magnetic field become different and, in general, 
diffusion can produce the inhomogeneous distribution of elements.

In this paper, we consider one more diffusion process that leads 
to formation of chemical inhomogeneities. This process is caused 
by a combined influence of the Hall effect and electric currents. 
Only fully ionized plasma is considered consisting 
electrons $e$, protons $p$, and small admixture of heavy ions $i$.
Generally, similar processes can occur in any system of charged
particles but they will be considered elsewhere. Using a simple 
model, we show that interaction 
of the electric current with impurities leads to their diffusion 
in the direction perpendicular to both the electric current and 
magnetic field. This type of diffusion can alter the distribution
of chemical elements and contribute to the formation of chemical 
spots even if the magnetic field is relatively weak and does 
not magnetize electrons ($x_e \ll 1$). We also argue that the 
current-driven diffusion in combination with the Hall effect 
can be the reason of the particular type of waves in which the 
impurity number density oscillates alone.

\section{Basic equations}

Consider a cylindrical plasma configuration with the magnetic
field parallel to the axis $z$, $\vec{B} = B(s) \vec{e}_{z}$; $(s, 
\varphi, z)$ and $(\vec{e}_s, \vec{e}_{\varphi}, \vec{e}_{z})$ are 
cylindrical coordinates and the corresponding unit vectors. 
Then, the electric current is 
\begin{equation}
j_{\varphi} = - (c/4 \pi) (d B/d s).  
\end{equation}
We suppose that $j_{\varphi} \rightarrow 0$ at large $s$ and,
hence, $B \rightarrow B_0$=const at $s \rightarrow \infty$.
Note that $B(s)$ cannot be an arbitrary function of $s$ because, 
generally, the magnetic configurations are unstable for
some dependences $B(s)$ (\cite{tay73,bu08a,bu08b}). The 
characteristic timescale of this instability is usually much 
shorter than the diffision timescale and, therefore, a formation 
of chemical structures in unstable magnetic configurations is 
impossible. Often, the magnetic field has a more complex topology 
than our simple model. However, this model describes correctly 
the main qualitative features of current-driven diffusion. In 
some cases, this model can even mimic the magnetic field in 
certain regions. For example, the field near the magnetic pole 
has a topology very close to our model \cite{uvr93}.    

We assume that plasma is fully ionized and consists of electrons 
$e$, protons $p$, and small admixture of heavy ions $i$. The 
number density of species $i$ is small and it does not influence 
the dynamics of plasma. Therefore, these ions can be treated as 
test particles interacting only with a background hydrogen plasma. 

The partial momentum equations in fully ionized plasma have 
been considered by a number of authors (\cite{brag65,urp81}). 
The paper \cite{urp81} deals mainly with the hydrogen-helium 
plasma. However, the derived equations can be applied for 
hydrogen plasma with a small admixture of any other ions if
their number density is small. If the mean hydrodynamic velocity 
is zero and only small diffusive velocities are non-vanishing, 
the partial momentum equation for the species $i$ reads
\begin{equation}
- \nabla p_i + Z_i e n_i \left( \vec{E} + 
\frac{\vec{V}_i}{c} 
\times \vec{B} \right) + \vec{R}_{i e} + \vec{R}_{i p} + 
\vec{F}_i
= 0 
\end{equation}
(\cite{brag65,vek87,urp81}), where $Z_i$ is the charge number of 
the species $i$, $p_i$ and $n_i$ are the partial pressure and number 
density, respectively, $\vec{V}_i$ is the velocity, and $\vec{E}$ 
is the electric field. The force $\vec{F}_i$ is the external force 
on species $i$; in astrophysical conditions, $\vec{F}_i$ is usually 
the sum of the gravitational and radiation forces. For the sake of 
simplicity, we neglect external forces in our simplified model. The 
forces $\vec{R}_{i e}$ and $\vec{R}_{i p}$ are caused by the interaction 
of ions $i$ with electrons and protons, respectively. The forces 
$\vec{R}_{i e}$ and $\vec{R}_{i p}$ are internal and their sum over 
all plasma components is zero in accordance with Newton's third law. 
Since diffusive velocities are small, we neglect the term proportional 
$(\vec{V}_i \cdot \nabla) \vec{V}_1$ in Eq.~(6).

If $n_i$ is small compared to the number density of protons, 
$\vec{R}_{ie}$ is given by      
\begin{equation}
\vec{R}_{ie} = - (Z_i^2 n_i / n) \vec{R}_{e}
\end{equation}
where $\vec{R}_{e}$ is the force acting on the electron gas 
\cite{urp81}. Since $n_i \ll n$, $\vec{R}_{e}$ is determined 
mainly by scattering of electrons on protons but scattering on 
ions $i$ gives a small contribution to $\vec{R}_{e}$. Therefore, 
we can use for $\vec{R}_{e}$ the expression for one component 
hydrogen plasma calculated by Braginskii \cite{brag65}. In our 
model of a cylindrical isothermal plasma, this expression reads
\begin{equation}
\vec{R}_{e} = - \alpha_{\perp} \vec{u} + \alpha_{\wedge} \vec{b}
\times \vec{u},
\end{equation}   
where $\vec{u} = - \vec{j}/en$ is the current velocity of 
electrons; $\vec{b} = \vec{B}/B$; $\alpha_{\perp}$ and 
$\alpha_{\wedge}$ are coefficients calculated by \cite{brag65}. 
The force (4) describes the standard friction caused by the 
relative motion of electrons and protons. Taking into account 
Eq.(1), we have 
\begin{equation}
\vec{u} = (c/4 \pi e n) (d B /d s) \vec{e}_{\varphi}.
\end{equation}

In this paper, we consider the current-driven diffusion in a 
relatively weak magnetic field that does not magnetize
electrons ($x_e \ll 1$). Substituting Eq.(4) into Eq.(3) and 
using coefficients $\alpha_{\perp}$ and $\alpha_{\wedge}$ from
\cite{brag65} with the accuracy in linear terms in $x_e$, we 
obtain 
\begin{equation}
R_{ie \varphi} \!\! = \! Z_i^2 n_i \left( \! 0.51\frac{m_e}{\tau_e} u 
\! \right), 
R_{ie s} \!= \!\! Z_i^2 n_i \left( 0.21 x \! \frac{m_e}{\tau_e} u 
\! \right). 
\end{equation}

If $T=$const, the friction force $\vec{R}_{ip}$ is proportional 
to the relative velocity of ions $i$ and protons, $\vec{R}_{ip} 
\propto (\vec{V}_p - \vec{V}_i)$. This force can be easily 
calculated in the case $A_i = m_i/m_p \gg 1$. Taking into account 
that the velocity of the background plasma is zero in our model, 
$\vec{R}_{ip}$ can be represented as \cite{urp81}
\begin{equation}
\vec{R}_{ip} = (0.42 m_i n_i Z^2_i / \tau_i) (-\vec{V}_i),
\end{equation} 
where $\tau_i = 3\sqrt{m_i} (k_B T)^{3/2} / 4 \sqrt{2 \pi} e^4 n 
\Lambda$; $\tau_{i}/ Z^2_i$ is the timescale of ion-proton 
scattering; we assume that $\Lambda$ is the same for all types 
of scattering.    

The momentun equation for the species $i$ (Eq.(2)) depends
on cylindrical components of the electric field, $E_s$ and 
$E_{\varphi}$. These components can be determined from the 
momentum equations for electrons and protons
\begin{eqnarray}
- \nabla (n k_B T) - e n \left( \vec{E} + \frac{\vec{u}}{c} \times
\vec{B} \right) + \vec{R}_e + \vec{F}_e = 0, \\
- \nabla (n k_B T) + e n \vec{E} - \vec{R}_e + \vec{F}_p = 0  
\end{eqnarray}
(\cite{brag65}). Taking into account the friction force $\vec{R}_e$  
(Eq.~(3)). we obtain with accuracy in linear terms in $x_e$ 
\begin{equation}
E_s \! = \! - \! \frac{uB}{2c} - \frac{1}{e} \! \left( \! 0.21 
\frac{m_e u}{\tau_e} x \! \right) , \; 
E_{\varphi} \! = \! - \! \frac{1}{e} \! \left( \! 0.51 
\frac{m_e u}{\tau_e}  \! \right).
\end{equation}
Substituting Eqs.(6), (7), and (10) into the $s$- and 
$\varphi$-components of Eq.(2), we arrive to the expression 
for a diffusion velocity, $\vec{V}_i$,
\begin{equation}
\vec{V}_i = V_{is} \vec{e}_s + V_{i \varphi} \vec{e}_{\varphi}, 
\;\;\;\;\; V_{is} = V_{n_i} + V_B,
\end{equation} 
where
\begin{equation}
V_{n_i} \!\!=\!\! - D \frac{d \ln n_i}{d s}, \;\;\; 
V_B \!\!=\!\! D_B \frac{d \ln B}{d s}, \;\;\;
V_{i \varphi} \!\!= \!\!D_{B \varphi} \frac{d B}{d s}; 
\end{equation}
$V_{n_i}$ is the velocities of ordinary diffusion and $V_B$ is 
the diffusion velocity caused by the electric current. The 
corresponding diffusion coefficients are
\begin{eqnarray}
D = \frac{2.4 c_i^2 \tau_i}{Z_i^2}, \;\;\; 
D_B = \frac{2.4 c_A^2 \tau_i}{Z_i A_i} (0.21 Z_i - 0.71), \\
D_{B \varphi} = 1.22 \sqrt{\frac{m_e}{m_i}} 
\frac{c (Z_i - 1)}{4 \pi en Z_i}. 
\end{eqnarray}
where $c_i^2 = k_B T/ m_i$ and $c_A^2 = B^2 / (4 \pi n m_p)$. 
Eqs.~(11)-(12) describe the drift of ions $i$ under the combined 
influence of $\nabla n_i$ and $\vec{j}$. 
The azimuthal drift velocity, $V_{i \varphi}$, is  
relevant to the current motion of electrons because heavy ions are
carried away by electrons in the direction of their motion. The 
radial velocity is caused mainly by the Hall effect. In the 
presence of electric currents, this effect produces a force that
is perpendicular to both the electric current (azimuthal) and 
magnetic field (vertical).

\section{Distribution of ions in the presence of electric currents}

In our model, the condition of hydrostatic equilibrium is given by
\begin{equation}
- \nabla p + \vec{j} \times \vec{B} / c = 0
\end{equation}
\cite{brag65}, where $p$ and $\rho$ are the pressure and density, 
respectively. Since the background plasma is hydrogen, $p \approx 
2n k_B T$ where $k_B$ is the Boltzmann constant. Integrating the 
$s$-component of Eq.~(15) and assuming that the temperature is 
constant, we obtain 
\begin{equation}
n = n_0 \left( 1 + \beta_0^{-1} - \beta^{-1} \right),
\end{equation}
where $\beta = 8 \pi p_0 / B^2$; $(p_0, n_0, T_0, \beta_0)$ are 
the values of $(p, n, T, \beta)$ at $s \rightarrow \infty$.

Consider the equilibrium distribution of elements. In equilibrium, 
we have $V_{i s} = 0$ and Eq.(11) yields 
\begin{equation}
\frac{d \ln n_i}{d s} =
\frac{D_B}{D} \frac{d \ln B}{d s}. 
\end{equation}
The term on the r.h.s. describes the effect of electric currents 
on the distribution of impurities. Note that this type of 
diffusion is driven namely by the electric current rather than 
an inhomogeneity of the magnetic field. Ocasionally, the conditions 
$d B/ ds \neq 0$ and $j \neq 0$ are equivalent in our simplified 
model. One has from Eq.~(15) 
\begin{equation}
\frac{d}{ds}(n k_B T) = - \frac{B}{8 \pi} \frac{d B}{d s}. 
\end{equation}  
Substituting Eq.~(18) into Eq.(17) and integrating, we obtain
\begin{equation}
\frac{n_i}{n_{i0}} = \left( \frac{n}{n_0} \right)^{\mu},
\end{equation}
where
\begin{equation}
\mu =  - 2 Z_i (0.21 Z_i - 0.71)
\end{equation}
and $n_{i0}$ is the value of $n_i$ at $s \rightarrow \infty$. 
Denoting the local abundance of the element $i$ as $\gamma_i = 
n_i/n$ and taking into account Eq.~(16), we have 
\begin{equation}
\frac{\gamma_i}{\gamma_{i0}} =
\left( \frac{n}{n_0} \right)^{\mu-1} = 
\left( 1 + \frac{1}{\beta_0} - \frac{1}{\beta} \right)^{\mu -1},
\end{equation}
where $\gamma_{i0} = n_{i0}/n_0$. Local abundances turn out to 
be flexible to the field strength and, particularly, this 
concerns the ions with large charge numbers. If other mechanisms 
of diffusion are neglidgible and the distribution of elements is 
basically current-driven, then the exponent $(\mu -1)$ can reach 
large negative values for elements with large $Z_i$ and, hence, 
produce strong abundance anormalies. For instance, $(\mu -1)$ is 
equal 1.16, -0.52, and -2.04 for $Z_i=$2, 3, and 4, respectively. 
Note that $(\mu -1)$ changes its sign as $Z_i$ increases: 
$(\mu -1) >0$ if $Z_i = 2$ but $(\mu -1)<0$ for $Z_i \geq 3$. 
Therefore, elements with $Z_i \geq 3$ are in deficit ($\gamma_i 
< \gamma_{i0}$) in the region with a weak magnetic field ($B < B_0$) 
but, on the contrary, these elements should be overabundant in 
the region where the magnetic field is stronger than $B_0$. 

Eq.~(21) describes the distribution of impurities in diffusive
equilibrium. The characteristic timescale to reach this 
equilibrium, $t_B$, can be estimated as
\begin{equation}
t_B \sim L / V_B \sim L^2 / D_B.
\end{equation} 
where $L$ is the magnetic lengthscale, $L = |d \ln B/ d s|^{-1}$. 
The characteristic timescale of baro-diffusion is given by the 
well-known expression
\begin{equation}
t_n \sim L / V_{ni} \sim L^2 / D.
\end{equation} 
Hence, the current-driven diffusion operates on a shorter 
timescale if $D_B > D$ or
\begin{equation}
\frac{c_A^2}{c_s^2} > Z_i^{-1} (0.21 Z_i - 0.71)^{-1},
\end{equation} 
where $c_s$ is the sound speed, $c_s^2 = k_B T/m_p$. Therefore,
the current-driven diffusion can be more efficient if the magnetic
pressure is greater than the gas pressure.

\section{Compositional waves}

The continuity equation for ions $i$ reads in our model \cite{urp81}
\begin{equation}
\frac{\partial n_i}{\partial t} + \frac{1}{s} \frac{\partial}{\partial s}
\left( s n_i V_{is} \right) + \frac{1}{s} \frac{\partial}{\partial
\varphi} (n_i V_{i \varphi} )  = 0.
\end{equation}
Consider the behaviour of small disturbances in the impurity 
number density, $n_i$, by making use of a linear analysis 
of Eq.~(25). Assume that plasma is in equilibrium in the 
unperturbed state. Since the number density of impurity 
is small, its influence on parameters of the basic state is 
negligible. We consider disturbances that do no depend 
on $z$. Denoting the disturbances of $n_i$ by $\delta n_i$ 
and linearizing Eq.(25), we obtain     
\begin{eqnarray}
\frac{\partial \delta n_i}{\partial t} - \frac{1}{s} 
\frac{\partial}{\partial s}
\left( s D \frac{\partial \delta n_i}{\partial s} - s \delta n_i
\frac{D_B}{B} \frac{dB}{ds} \right) +
\nonumber \\
\frac{1}{s} \frac{\partial}{\partial \varphi}
\left( \delta n_i D_{B \varphi} \frac{d B}{d s} \right)= 0.
\end{eqnarray}
We consider disturbances with the wavelength shorter than the 
lengthscale of $B$. In this case, we can use 
the so called local approximation and assume that disturbances
are $\propto \exp(-i k s -M \varphi)$ where $k$ is the wavevector,
$ks \gg 1$, and $M$ is the azimuthal wavenumber. Since the basic 
state does not depend on $t$, $\delta n_i$ can be represented 
as $\delta n_i \propto \exp(i\omega t - i k s -i M \varphi)$ where 
$\omega$ should be calculated from the dispersion equation.
We consider two particular cases of the compositional waves,
$M=0$ and $M \gg ks$.

{\it Cylindrical waves with $M=0$.}
Substituting $\delta n_i$ into Eq.~(26), we obtain
the dispersion equation for $M=0$ 
\begin{equation}
i \omega \!=\! - \omega_{R} \! + \! i \omega_{B}, \; \omega_{R} \!=
\!D k^2, \; \omega_{B} \!= \! k D_B (d \ln B / ds).
\end{equation}
This dispersion equation describes cylindrical waves in which 
only the number density of impurity oscillates. The quantity 
$\omega_{R}$ describes decay of waves with the 
characteristic timescale $\sim (D k^2)^{-1}$ typical for a
standard diffusion. The frequency $\omega_{B}$ describes
oscillations of impurities caused by the combined action of
electric current and the Hall effect. 
The frequency is non-vanishing only in the 
presence of electric currents since $dB/ds = - (4 \pi/cB) j_{\varphi}$.
Therefore, such waves cannot exist in multicomponent 
current-free plasma even if the magnetic field is sufficiently
strong to magnetize plasma. Note that the frequency 
can be of any sign but $\omega_{R}$ is always
positive. The compositional waves are aperiodic if $\omega_{R} >
|\omega_{B}|$ and oscillatory if $|\omega_{B}| > \omega_{R}$. This
condition is equivalent to  
\begin{equation}
c_A^2/c_s^2 > Z_i^{-1} |0.21 Z_i - 0.71|^{-1} kL,
\end{equation}
where $c_s$ is the sound speed, $c_s^2 = k_B T/m_p$.
The compositional waves become oscillatory if 
the field is strong and the magnetic pressure is
substantially greater than the gas pressure. The frequency 
is higher in the region where the magnetic
field has a strong gradient. The order of magnitude
estimate of $\omega_{I}$ is 
\begin{equation}
\omega_{I} \sim k c_A (1 / Z_i A_i) (c_A / c_i)(l_i / L),
\end{equation}
where $l_i = c_i \tau_i$ is the mean free-path of ions $i$. 
Note that different impurities oscillate with different 
frequences.

{\it Non-axisymmetric waves with $M \gg ks$.} In this case,
the dispersion equation reads
\begin{equation}
i \omega \!= \! - \omega_{R} \!+\! i \omega_{B \varphi}, \;\; 
\omega_{B \varphi} \!= \! (M/s) B D_{B \varphi} (d \ln B / ds).
\end{equation}
Like cylindrical waves, the non-axisymmetric compositional 
waves exist only in the presence of electric currents.
Non-axisymmetric waves rotate around the cylindric axis
with the frequency $\omega_{B \varphi}$ and decay slowly on 
the diffusion timescale $\sim \omega_R^{-1}$. The frequency
of such waves is typically higher than that of cylindrical 
waves. One can estimate the ratio of these frequencies as
\begin{equation}
(\omega_{B \varphi} / \omega_B) \sim (B D_{B \varphi} / D_B)
\sim (1 / A_i x_e) (M / ks).
\end{equation}
Since we consider only weak magnetic fields 
($x_e \ll 1$), the period of non-axisymmetric waves
is shorter for waves with $M > A_i x_e (ks)$. The ratio 
of the diffusion timescale and period of non-axisymmetric waves
is 
\begin{equation}
(\omega_{B \varphi} / \omega_R) \sim (1 / x_e) 
(c_A^2 / c_s^2) (Z_i / A_i) (1 / kL)
\end{equation} 
and it can be large. Hence, these waves can be oscillatory.

\section{Discussion}

We have considered diffusion of heavy ions under 
the influence of the current-driven mechanism. This mechanism is 
well known in plasma physics (see \cite{vek87} for review) and can
be responsible for many phenomena in laboratory devices. Generally, the 
velocity of such diffusion can be comparable to or even greater 
than that caused by other diffusion mechanisms. It turns out that
the electric currents lead to the formation of chemical inhomogeneities
(spots) in multicomponent plasma. The chemical composition in such
spots can differ substantially from the average one (see Eq.~(21)).
The efficiency of the considered diffusion mechanism depends on a 
strength of the magnetic field. This mechanism can 
operate even if the magnetic field is relatively weak whereas other 
diffusion mechanisms require a substantially stronger magnetic 
field. For example, some impurities in astrophysical conditions can 
drift under the influence of a radiation force \cite{mich70,mich76}. 
This occurs in bright stars with a high surface temperature. However,
such diffusion is influenced by the magnetic field only if it is
sufficiently strong and can magnetize ions.  

The current-driven diffusion is relevant to the Hall effect and,
therefore, it leads to a drift of ions in the direction perpendicular 
to both the magnetic field and electric current. As a result, a 
distribution of chemical elements in plasma depends essentially on 
the geometry of the magnetic field and electric current. In our
simple model of the magnetic field, chemical inhomogeneities have a
cylindrical form but, generally, their distribution can be much 
more complicated. A cylindrical geometry is suitable for various 
experimental designs and often used for numerical modelling
(see \cite{vek87} for a review).  
Chemical inhomogeneities can manifest themself by emission in 
spectral lines and a non-uniform distribution of the 
temperature. Note that the considered mechanism of diffusion can 
be important not only in plasma but in some conductive fluids if 
the magnetic field is sufficiently strong.  

Our study reveals that a particular type of waves may exist in
multicomponent plasma in the presence of electric currents. These 
waves are slowly decaying and characterized by oscillations of the 
impurity number density alone. They exist only if the magnetic 
pressure is greater than the gas pressure. Such condition is fulfilled
in many laboratory experiments. The frequency of compositional 
waves turns out to be different for different impurities. These
waves should manifest itself by oscillations in spectrum.    
Compositional waves can occur in both astrophysical and laboratory 
plasmas but their frequency is essentially higher in laboratory
conditions. For example, if $B \sim 10^5$ G, $n 
\sim 10^{15}$ cm$^{-3}$, $T \sim 10^6$K, and $L \sim \lambda \sim 
10^2$ cm, then the period of compositional waves is $\sim 10^{-8}$ s. 
Note that this is only the order of magnitude estimate but 
frequencies of various impurities may differ essentially since the 
period of compositional waves depend on the sort of heavy ions.  \\

Usually, 
diffusion processes play an important role in plasma if hydrodynamic
motions are very slow. In some cases, however, chemical spots can
be formed even in flows with a relatively large velocity but with
some particular topology (for example, a rotating flow). This can 
occur usually in laminar flows. Unfortunately, such flows often are 
unstable in magnetized plasma. This is particulary concerned by 
flows with a large Hall parameter since hydrodynamic motions in 
such plasma typically are unstable even in the presence of a 
very small shear (\cite{mik09,bej11,urp05,bon06}). As a
result, a formation of the chemical spots is unlikely if there are 
hydrodynamic motions even with a relatively weak shear.


{\it Acknowledgements}. The author thanks the Russian
Academy of Sciences for financial support under the 
programme OFN-15.



\end{document}